\documentclass[11pt]{article}

\usepackage{graphicx}

\newcommand{\p}{$\;\;$}

\def\SDiff{S\kern-1.5pt di \kern-1pt f \kern-1.5pt f}

\begin{document}

\begin{center}

{\bf 
Is there a common origin for the WMAP low multipole and
for the ellipticity in BOOMERanG CMB maps?
}

\end{center}

\vspace{0.2in}

\noindent V.G.Gurzadyan$^{1,2}$, P.A.R.Ade$^3$, P.
de Bernardis$^4$, C.L.Bianco$^2$, J.J.Bock$^5$, A.Boscaleri$^6$,
B. P. Crill$^7$, G. De Troia$^4$, 
E.Hivon$^8$, V.V.Hristov$^{7}$, A.L.Kashin$^1$, A.E.Lange$^7$,
S.Masi$^4$, P.D.Mauskopf$^3$, T.Montroy$^{9}$, P. Natoli$^{10}$,
C.B.Netterfield$^{11}$, E.Pascale$^6$, F.Piacentini$^4$,
G.Polenta$^4$, J.Ruhl$^{9}$, G.Yegorian$^{1}$

\vspace{0.2in}

$^1$ Yerevan Physics Institute (Armenia)

$^2$ ICRA, Dipartimento di Fisica, University La Sapienza, Roma
(Italy)

$^3$ Department of Physics and Astronomy, Cardiff (UK)

$^4$ Dipartimento di Fisica, University La Sapienza, Roma (Italy)

$^5$ JPL, Pasadena (USA)

$^6$ IROE-CNR, Firenze (Italy)

$^7$ Caltech, Pasadena (USA)

$^8$ IPAC, Pasadena (USA)

$^9$ Department of Physics, U.C. Santa Barbara (USA)

$^{10}$ Dipartimento di Fisica, Tor Vergata, Roma, (Italy)

$^{11}$ Department of Physics, University of Toronto (Canada)

\vspace{0.2in}

{\bf Abstract} - We have measured the ellipticity of several degree scale anisotropies in the BOOMERanG maps of the Cosmic Microwave Background (CMB) at 150 GHz. The average ellipticity is around 2.6-2.7. The biases of the estimator of the ellipticity and for the noise are small in this case.  Large spot elongation had been detected also for COBE-DMR maps. If this effect is due to geodesic mixing, it would indicate a non precisely zero curvature of the Universe which is among the discussed reasons of the WMAP low multipole anomaly. Both effects are related to the diameter of the Universe: the geodesics mixing through hyperbolic geometry, low multipoles through boundary conditions. This common reason can also be related with
the cosmological constant: the modes of vacuum fluctuations conditioned by the boundary conditions lead to a value of the cosmological constant being in remarkable agreement with the supernovae observations.

\section{Introduction}

The BOOMERanG 1998 experiment provided crucial information on the acoustic peaks of the power spectrum of CMB and hence on the conditions at the last scattering epoch and cosmological parameters \cite{Bern1}, \cite{Bern2}. The Boomerang maps were compared with WMAP data \cite{Bern3}. 

The 150 GHz Boomerang maps were also analyzed to trace the geometrical shapes, namely, the ellipticity of the anisotropies \cite{ellipse}. That analysis was motivated by similar analysis of the COBE-DMR data, where a signature of elongation of anisotropies had been detected \cite{GT}. The latter analysis itself was motivated by 
the effect of geodesic mixing \cite{GK1}, which can result in elongation of CMB anisotropies independent on the temperature threshold. More generally, however, such analysis implies the estimation of Lyapunov exponents of any dynamical process which might lead to such distortion whatever its physical nature can be, as often used for example in the analysis of hydrodynamical experiments.

The study of the Boomerang maps shows the existence of a temperature interval where the elongation was independent on the threshold. The mean elongation yielded 2.2-2.3 for hot and cold areas (spots). The orientation of the
elongations did not show any preferred direction. In the present paper we report the results of the largest anisotropy areas, namely containing more than 60 and 100 pixels.
The importance of such analysis is related to the fact that the biases of the estimator and of the noise (which were studied by simulated maps and which could not be ignored for smaller,
several pixel areas) are negligible for larger areas. 

The low quadrupole detected by WMAP \cite{Ben03} led to a hot debate on the cosmological model: the curvature and/or topology of the Universe among the disputed reasons (e.g. \cite{Aur03,Efs03,Ell03,Lum03}).
It is remarkable that the ellipticity analysis even though carried out in a relatively small region of the sky (about 1 per cent) nevertheless might have the ability to detect such a large scale effect. Indeed, if the elongation is due to geodesic mixing than it has to refer to the hyperbolicity of the Universe and hence the non zero negative curvature. About the topology nothing can be claimed, since the given curvature can correspond to various both open and closed topologies. 

\section{Ellipticity in CMB maps and geodesic mixing}

We briefly mention the idea of geodesic mixing as an effect which might produce the observed elongations. Photon beams while moving in hyperbolic spaces have exponentially deviating trajectories are described by the equation of geodesics deviation
\begin{equation}
  \frac{d^2{\bf n}}{d\lambda^2}+k{\bf n}=0,
\end{equation}
where $n$ is the deviation vector. The solution of this equation is
\begin{equation}
{\bf n (\lambda)}={\bf n(0)}cosh \chi \lambda + \dot{{\bf n}}(0)sinh  \chi \lambda
\end{equation}
when $k=-1$. On the other hand, it is proved that the geodesic flows
in hyperbolic spaces (locally if the space is not compact) behave as Anosov systems, \cite{Anosov} i.e. dynamical systems with strong statistical properties, which can be described by proper reparameterisation of the affine parameter $\lambda$ for null geodesics \cite{LMP}. Then for geodesic flows of spaces of Robertson-Walker metric $\chi$  is the Kolmogorov-Sinai entropy and is given by the diameter of the Universe \cite{GK1}
\begin{equation}
\chi=1/a,
\end{equation}
with the decay of time correlation function of the geodesic flow
\begin{equation}
|b(t)|\leq |b(0)| e^{-c \chi \lambda}, c > 0.
\end{equation}
In the context of freely propagating CMB photon beams this implies several observational effects such as the isotropization of the photon beams, i.e. damping of the degree of anisotropy by time, as well as to the distortion in the 
maps \cite{GK1}.\footnote{Geodesics or chaotic mixing is not however an alternative to inflation as mentioned e.g. in \cite{silk} but can act as an alternative tool to trace the geometry.}
The geodesic mixing is a statistical effect
independent on the conditions on the last scattering surface, and
is distinguished by threshold independence and random
obliquities of the elongated areas. 
The measured Lyapunov exponents and Kolmogorov-Sinai entropy can describe also other processes occuring here. More informative descriptor for more accurate maps is the Kolmogorov complexity \cite{G}.

\section{BOOMERanG's Data}

BOOMERanG is a millimetric telescope with bolometric detectors on
a balloon borne platform \cite{Bern2}, \cite{Pia}, \cite{Masi3}, \cite{Crill}. 
It was flown in 1998/99 and produced wide (4$\%$
of the sky), high resolution ($\sim 10'$) maps of the microwave
sky (90 to 410 GHz) \cite{Bern1}. In the maps the structure is
resolved with high signal to noise ratio, and hundreds of
degree-scale hot and cold areas are evident. The rms temperature
fluctuation of these areas is $\sim 80 \mu K$. The detected
fluctuations are spectrally consistent with the derivative of a
2.735 K blackbody. Masi et al. \cite{Masi3} have shown that
contamination from local foregrounds is negligible in the maps at
90, 150 and 240 GHz, and that the 410 GHz channel is a good
monitor for dust emission.

In our study we used two maps from independent detectors at 150 GHz. 
These maps have been
obtained from the time ordered data using an iterative procedure
\cite{Nato01}, which properly takes into account the system noise
and produces a maximum likelihood map. The largest structures
(scales larger than 10$^o$) are removed in this procedure, to
avoid the effects of instrument drifts and 1/f noise.
The two input maps, A and B, included 33111 pixels, each of ~7
 arcmin in linear size, in a high Galactic latitude region covering
about 1 \% of the sky, with coordinates $RA > 70^{\circ}$,
$-55^{\circ} < dec < -35^{\circ}$ and $b < -20^o$ \cite{Bern1}.
The first map (A) has been obtained from the data of the B150A
detector, while the second map (B) was obtained by averaging the
maps from detectors B150A1 and B150A2. In this way we obtained two
maps with similar noise per pixel. The signal (CMB) to noise ratio
per pixel is of the order of 1 for our 7' pixels. The sum (A+B)
is shown in Fig 1. There are three AGN with significant flux in the
maps. All the areas including
these AGNs have been excluded from the analysis.

\begin{figure}[htp]
\begin{center}
\includegraphics[height=8cm,width=8cm]{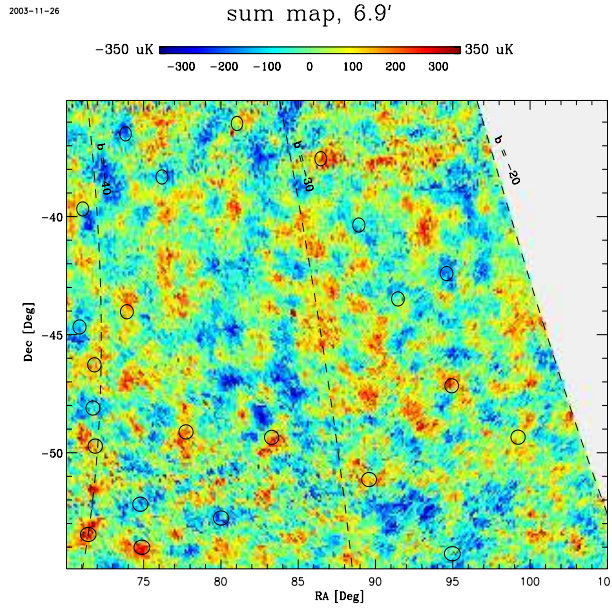}
\caption{Sum (A+B) map obtained from three independent measurement
channels at 150 GHz: A+B= B150A+(B150A1+B150A2)/2. The pixel size
is 6.9 arcmin (Healpix npix=512). The measurement units ($\mu K$)
refer to thermodynamic temperature fluctuations of a 2.73K
Blackbody. The circles locate the anisotropy spots with more than
100 pixels detected at a threshold level of $\pm 100 \mu K$ (see
Tables 1a and 1b).
 }
\end{center}
\end{figure}

\section{Analysis}

The algorithms of the study of excursion sets (areas) in the maps, i.e. 
pixel sets with temperature equal and higher than a given temperature threshold (lower, for negative
thresholds) in the BOOMERanG maps are described in details in \cite{ellipse}.
The sensitivity of the procedures has been checked not only varying various input parameters but also via simulated maps.  This enables to evaluate the biases of the ellipticity estimator (up to 0.4) and the bias due to the noise (up to 0.3). 
It was clear however that for larger areas these biases are much smaller, less than 0.1. 
Therefore the study of large areas (i.e. several degree-scale) containing over 60 or 100 pixels has a 
particular interest. The results are shown in Fig.2 where the existence of threshold independent ellipticity 
is evident. The obliquities of the areas again do not show a preferred direction (Fig.3).

Tables 1ab contain the data for the hot and cold anisotropy areas of more than 100 pixels at
thresholds $\pm 100 \mu K$.

\section{Conclusions}

The analysis of degree scale areas in the Boomerang CMB sky map reveals threshold
independent ellipticity of around 2.6-2.7 in the meaningful
interval of the temperature thresholds. 
Ellipticity for large scale areas had been found also in COBE-DMR maps \cite{GT}. 
Similar analysis of the WMAP data is underway.
The angular distribution of the areas is random.
So, the effect exists not only for scales smaller than the horizon at the last scattering surface 
but also on larger scales. Thus it cannot be due to the physical effects at
the last scattering surface: it has to arise after. Both effects, the geodesics mixing and the low multipoles are related to the diameter of the Universe: the first one via boundary conditions, the second one via hyperbolic geometry. 
Thus, if the detected ellipticity is due to geodesic mixing than this would indicate the non precisely zero curvature which might be the reason also of the low multipole behavior observed by WMAP. 

The boundary conditions and the geometry of the Universe if taken into account for the modes of the vacuum fluctuations contributing into the cosmological constant predict a value being in remarkable agreement with the supernovae observations \cite{Lambda}.

\begin{figure}[htp]
\begin{center}
\includegraphics[height=8cm,width=14cm]{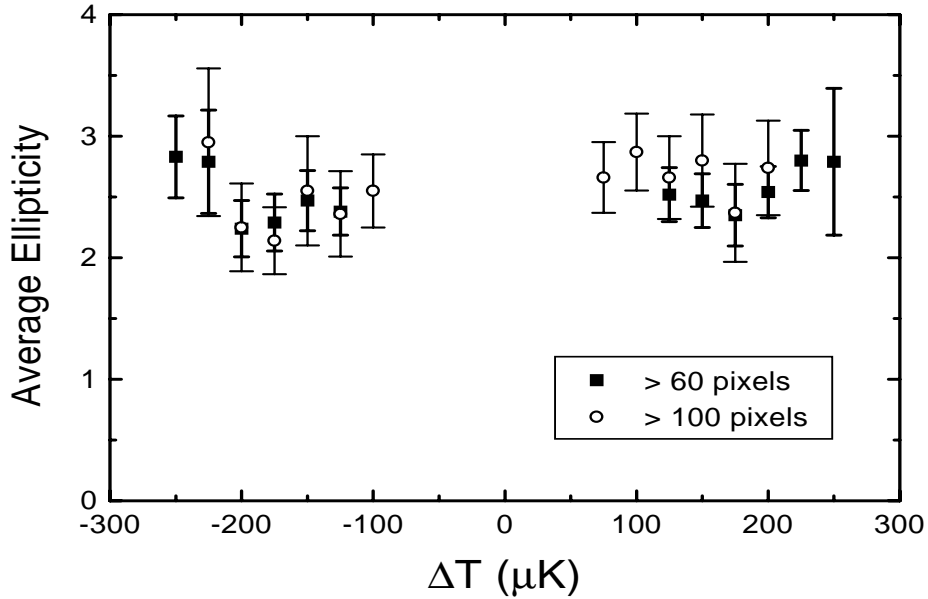}
\caption{Ellipticity vs temperature threshold (in $\mu K$)  for
the anisotropy areas containing more than 60 (squares) and 100 pixels (circles)
in the sum map of three independent 150 GHz channels.  }
\end{center}
\end{figure}
\begin{figure}[htp]
\begin{center}
\includegraphics[height=6cm,width=9cm]{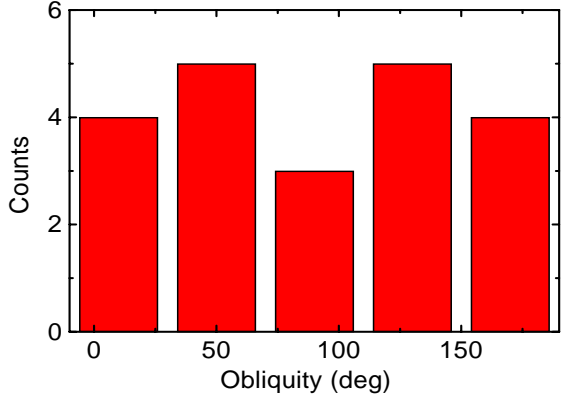}
\caption{Obliquities of hot and cold anisotropy areas of temperature threshold 150 and 100 $\mu K$, respectively, 
containing more than 100 pixels of the 150 GHz sum map.  }
\end{center}
\end{figure}

\begin{table}
{\renewcommand{\baselinestretch}{1.2}
\renewcommand{\tabcolsep}{5.4mm}

\begin{center}

                             Table 1a.
\end{center}
\begin{center}

{\small
      Threshold: 100 $\mu K$ \\[3mm]
\begin{tabular}{lcccc}
\hline
\hline
Area  &\multicolumn{2}{c}{Coordinates, degree}& Number   & Ellipti-\\

\p No &   l        &       b                & of pixels& city    \\
\hline
\p  h1&  243.3&   -40.7&  \p133&   3.80\\ %
\p  h2&  252.0&   -40.3&  \p177&     2.33\\ %
\p  h3&  256.5&   -40.1&  \p111&   3.69\\ %
\p  h4&  261.4&   -40.0&  \p113&   1.73\\ %
\p  h5&  249.1&   -38.8&  \p123&   3.33\\ %
\p  h6&  261.8&   -37.9&  \p134&   2.96\\ %
\p  h7&  255.6&   -36.3&   \p193&   2.74\\ %
\p  h8&  240.2&   -32.4&  \p103&   2.85\\ %
\p  h9&  256.1&   -32.7&   \p105&   2.13\\ %
\p h10&  258.7&   -28.9&  \p125&   3.96\\ %
\p h11&  243.0&   -28.5&  \p189&   2.19\\ %
\p h12&  255.1&   -24.7&  \p139&   2.57\\ %
\p h13&  258.2&   -22.5&  \p123&   2.98\\ %
\hline
\end{tabular}}
\end{center}}
\end{table}
\begin{table}
{\renewcommand{\baselinestretch}{1.2}
\renewcommand{\tabcolsep}{5.4mm}
\begin{center}

                             Table 1b
\end{center}
\begin{center}
{\small
      Threshold: -100 $\mu K$ \\[3mm]
\begin{tabular}{lcccc}

\hline \hline
Area &\multicolumn{2}{c}{Coordinates, degree}& Number   & Ellipti-\\
\p No  &   l        &        b               & of pixels& city    \\
\hline
\p  c1&  249.9&   -41.0&     113&  5.05\\ %
\p  c2&  254.4&   -40.3&     109&  1.88\\ %
\p  c3&  259.5&   -38.1&     133&  1.68\\ %
\p  c4&  239.5&   -38.2&     118&  2.75\\ %
\p  c5&  242.1&   -36.6&     102&  1.60\\ %
\p  c6&  260.1&   -34.9&     136&  2.19\\ %
\p  c7&  246.6&   -27.3&     189&  2.41\\ %
\p  c8&  262.8&   -26.2&     168&  2.00\\ %
\p  c9&  250.5&   -26.2&     166&  3.90\\ %
\p c10&  250.0&   -23.7&     101&  2.05\\ %
\hline
\end{tabular}}
\end{center}}
\end{table}

\end{document}